\documentstyle[aps,epsf,twocolumn]{revtex}
\begin{document}
\draft
\title{Parity Effect in the Resonant Tunneling}
\author{Seiji MIYASHITA, Keiji SAITO, and Hiroto KOBAYASHI}
\address{
Department of Applied Physics, Graduate School of Engineering,  \\
University of Tokyo, Bunkyo-ku, Tokyo 113-8656, Japan}
\date{\today }
\maketitle
\begin{abstract}
A mechanism of the parity effect in the thermally assisted resonant
tunneling is proposed in the view point of nonadiabatic transitions
of thermally excited states. In this mechanism, alternating enhancement
of the relaxation is naturally understood 
as a general property of quantum relaxation
of uniaxial magnets at finite temperatures where appreciable populations
are pumped up to excited states. It is also found that the 
enhanced sequence depends on the sweeping rate of the field. 
\end{abstract}
\noindent
\pacs{PACS number: 75.40.Gb,76.20.+q}
As to the relaxation of metastable magnetization of 
uniaxial nanoscale molecular magnets such as Mn$_{12}$
and Fe$_8$, the resonant tunneling phenomena have been
paid attention and various interesting properties of the
phenomena have been reported\cite{BB1,TART,exp1,exp2,exp5,expP,Fe1,Fe2,Fe3,Fe4}.
The key mechanism of the relaxation comes from their discrete energy structure due to a
finite number of degrees of freedom. 
The eigenvalues of the system are functions of the parameters of the system,
such as an external field. 
If we change a parameter infinitesimally small, 
then the system changes adiabatically, i.e., if the
system is initially in the ground state, then it stays in the ground state
of the system with the current value of the parameter. 
On the other hand,
if the changing rate is finite, the system cannot completely follow the change 
of parameter and then so-called nonadiabatic transition occurs.
For example in uniaxial magnetic systems, if we sweep 
the field very slowly from parallel to antiparallel to the initial 
magnetization,
the magnetization adiabatically follows the field and reverses its direction. 
This change of magnetization 
corresponds to the tunneling (the adiabatic transition).
If the sweeping rate is fast, then the magnetization only partially changes
(the nonadiabatic transition).

In uniaxial magnets, quantum fluctuation is relevant 
only at avoided level crossing points and changes of magnetization
only occur at those points.
The sweeping rate dependence of the probability of staying in the
original state in this type of nonadiabatic transition has been given by 
Landau \cite{Landau}, Zener \cite{Zener}, and St\"ukelberg \cite{St} (LZS).
We have studied changes of magnetization in a sweeping field from the view
point of the LZS mechanism\cite{miya95,miya96,DMSGG} and proposed to obtain
the tunneling gap from the magnetization change 
in a sweeping field\cite{miya96} and also
explained the step-like magnetization process as a characteristic feature of
nanoscale magnets\cite{DMSGG}.
The LZS mechanism is pure quantum mechanical and it is independent of the 
temperature. However in experiments, 
strong temperature dependences have been observed, which brought an idea
``thermally assisted resonant tunneling''\cite{TART,exp2}.
In order to explain this temperature dependence,
various theoretical attempts have been done\cite{GC97,LBF98,FRVGS98,KN98,DZ97,G97}.

The sweeping rate dependences of magnetization process 
have been also observed in experiments
at very low temperatures in Mn$_{12}$\cite{expP} 
and also in Fe$_{8}$\cite{Fe3,Fe4}. There data do not depend on the temperature
any more. We have pointed out that even in such cases 
there is still inevitable effect 
of the environment\cite{SMD}, and that we could nevertheless 
estimate the pure quantum transition probability.

Recently new aspects of the resonant tunneling have been reported\cite{BB1},
e.g., the parity effect of the resonant tunneling where amount of
relaxation changes at the resonant points alternately, 
and $\sqrt{t}$-dependence of the initial relaxation 
of the magnetization at the resonant points\cite{PS0}. 
In the present Letter,
we would like to propose a mechanism of the parity effect as 
a universal property of the thermally assisted resonant tunneling.

In order to investigate characteristics of temperature dependence of the
resonant tunneling, we have constructed an equation of motion of
the density matrix where effects of
a thermal bath are taken into account
(the quantum master equation\cite{QME}):
\begin{eqnarray}
\frac{\partial\rho(t)}{\partial t} &=& 
-i \left[{\cal H},\rho(t)\right] 
-\lambda
\left( \left[X,R\rho(t)\right] + \left[X,R\rho(t)\right]^{\dag} \right) , 
\label{CTTR}  
\end{eqnarray}
where $ {\cal H} $ is the Hamiltonian, $\rho (t)$ is 
the density matrix of the system and $X$ is a system operator
through which the system and the bath couple with the constant 
$\lambda$.
Here we set $\hbar $ to be unity.
The first term of the right-hand side
describes the pure quantum dynamics of the system while the second term
represents effects of environments at a temperature $T(=\beta^{-1})$.
There $R$ is defined as follows: 
\begin{eqnarray}
\langle k | R  | m \rangle &=& 
\zeta (E_{k} - E_{m})
n_{\beta} ( E_{k} - E_{m} )  
\langle k | X  | m \rangle , \nonumber \\
\zeta (\omega ) &=& I(\omega ) -I(- \omega) , 
\quad {\rm and} \quad
n_{\beta}( \omega ) =  ({e^{\beta\omega} -1 })^{-1}, \nonumber 
\end{eqnarray}
where
$| k \rangle $ and $| m \rangle $ represent the eigenstates of ${\cal H}$ 
with the eigenenergies $E_{k}$ and $E_{m}$, respectively.
Here we adopt a thermal bath which consists of an infinite number of
bosons ${\cal H}_{\rm B}= \sum_{\omega} \omega b_{\omega}^{\dagger}b_{\omega}$,
where $b_{\omega}$ and $b_{\omega}^{\dagger}$ are the 
annihilation and creation boson operators of the frequency $\omega$.
We adopt the spectral density of the boson bath $I(\omega )$ in the form
$I (\omega ) = I_{0} \omega^2 $, % is adopted\cite{GSI88}.
which is associated with phonon reservoir.
As to the interaction between the system and the bath we adopt a form
$X \sum_{\omega }(b_{\omega}+b_{\omega}^{\dagger})$.% taking $X=S_x+S_z$.

 As a more relevant source of a noise at very low 
temperatures,  we may consider the dipole field from other molecules or$/$and
the hyperfine interaction from nuclear spins\cite{PS96}.
For such noises we have to take into account another contribution to $R$.
In this sense the bath treating here does not represent the
experimental situation very appropriately.  
In the present Letter, however, we discuss only general natures 
which do not depend on the detail of the thermal bath. 

Equation (\ref{CTTR}) with $X=S_x+S_z$  was used to study the 
aforementioned inevitable effects of environments in  
the magnetization process under a sweeping field 
at a very low temperature \cite{SMD}.
%In such a low temperature, no temperature dependence is observed 
%in the experiment\cite{expP}, but a simple application of the LZS process fails to explain the
%phenomena. 
%Effects of environments has been explained as an effect of dissipative
%environments\cite{SMD}. 
%There we found that the effect of the quantum transition 
%(LZS process) and the dissipative effect can be separated, so that 
%we can obtain the quantum mechanical transition rate from the experimental 
%data.  
For the process the existence of interaction is essential
but detailed nature of the mechanism of the dissipation is not  
important. Thus we could discuss the property of the process as
a universal property of the relaxation on nanoscale magnets.
 
When the temperature goes up and the effect of the bath increases, 
the dissipative process becomes
to depend on specific features of the bath and the coupling.
Thus it becomes difficult to treat relaxation without specifying
nature of the bath.
However, in the present Letter, 
we will point out that the aforementioned parity effect is 
a universal property of the resonant tunneling at finite temperatures, 
which is independent of detailed nature
of the bath.  
%Thus here we use the same equation as that used in the previous paper. 

Let us consider a general model of an $S=10$ uniaxial magnet
in an external field:
\begin{equation}
{\cal H}=-DS_z^2-\Gamma S_x-H(t)S_z+Q,
\label{ham}
\end{equation} 
where $Q$ represents extra terms such as $(S^+)^4+(S^-)^4$, etc.
We will propose a mechanism of the parity effect as a general property 
of uniaxial magnets. Thus we put $Q=0$. 

In Fig.\ 1(a) we show a magnetization process $M(t) 
= \langle S_z \rangle / S $ for the case with
$T=1.0$, $\lambda=0.00005$, $\Gamma=-0.45$. 
We sweep the magnetic field $H(t)= ct-H_{0}$ with $c=0.0001$ 
and $H_{0}= 0.3$. 
In Fig.\ 1(b), 
the derivative $dM/dH (=c^{-1} dM/dt)$ is also shown.  
In these figures, we find an alternate change of the amount of changes of 
the magnetization clearly. 
\epsfxsize=9.0cm \epsfysize=6.0cm \epsfbox{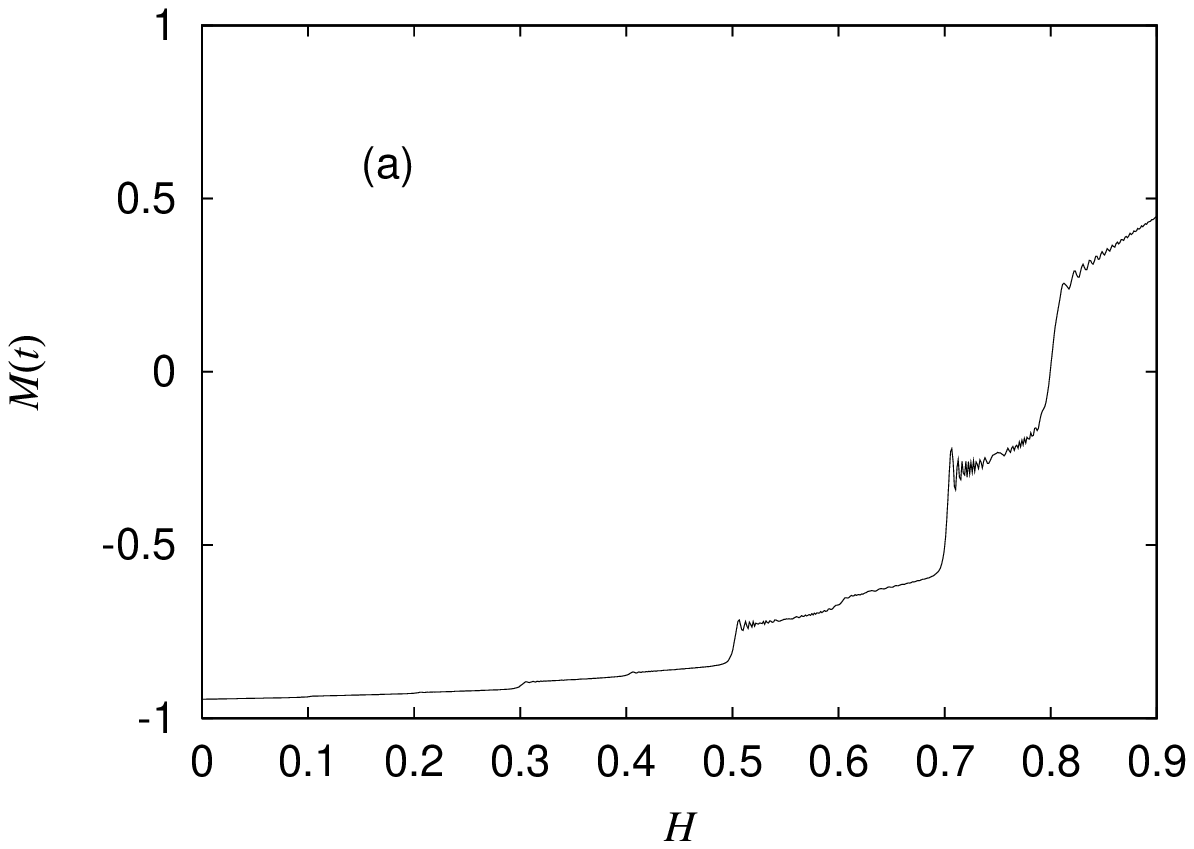} 
\begin{figure}
\noindent
\epsfxsize=9.0cm \epsfysize=6.0cm \epsfbox{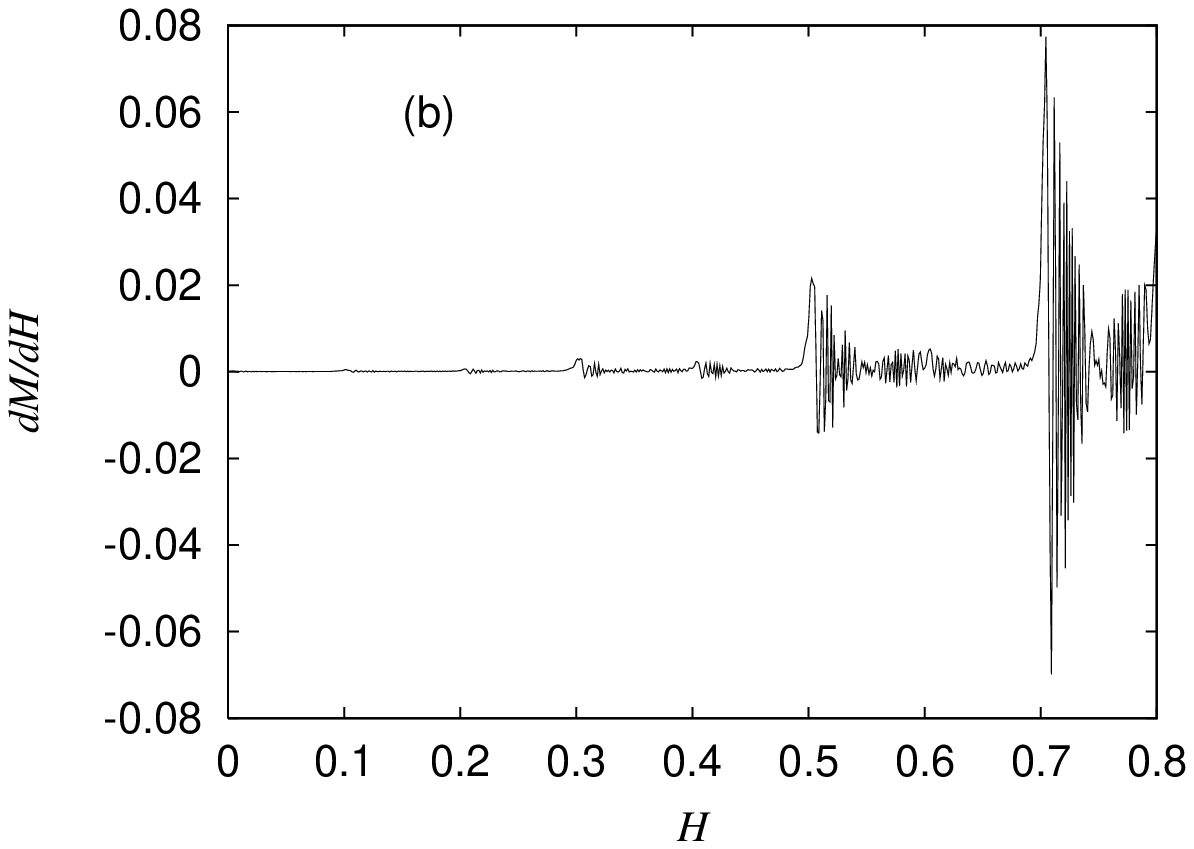}
\caption{(a) Magnetization process and (b) $dM/dH$ for $c=0.0001$.}
\end{figure}
\begin{figure}
\noindent
\epsfxsize=9.0cm \epsfysize=6.0cm \epsfbox{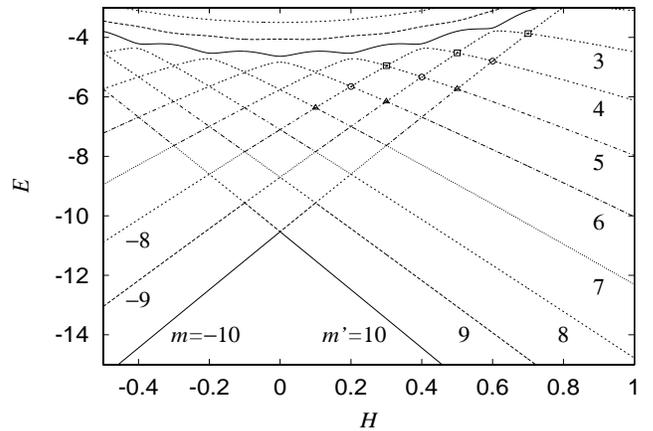} \\
\caption{Energy structure as a function of the external field of the model
(\ref{ham}) with $D=0.1,\Gamma=0.45$.}
\end{figure}

In Fig.\ 2, energy levels as a function of the field
are shown. 
Here we see straight lines along which the magnetization is approximately given by
$m$ or $m'$. These lines denoting energy levels for the diagonal parts of the 
Hamiltonian 
are called diabatic states.
At each crossing point, a small energy gap is created by the off-diagonal
terms and so-called avoided level crossing is formed, where
large enhancement of relaxation occurs (resonant tunneling).
The energy gaps of avoided level crossing
points are listed in Table I, where we denote avoided level crossing points
of levels of $m$ and $m'$ by $(m,m')$.
There we find that the energy gaps at the same
horizontal level in Fig. 2 (denoted by the same symbols) 
are about the same. This is easily
understood from the fact that the gap is of order $\Gamma^{|m-m'|}$ 
where $m$ and $m'$ are the magnetizations of the crossing levels for 
$\Gamma=0$ \cite{CG}.

In order to see what processes are going, we show, in Fig.\ 3, 
the time evolution of distribution of occupation probabilities at the $i$th 
level, which is expressed by $\langle i|\rho|i\rangle$.

\begin{figure}
\noindent
\epsfxsize=9.0cm \epsfysize=6.0cm \epsfbox{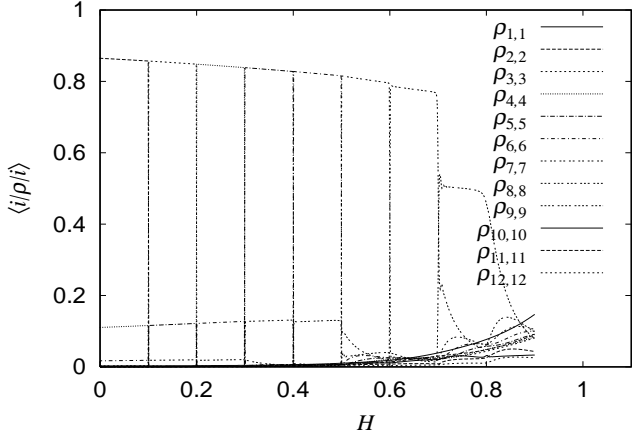}
\caption{Time evolution of the population $\langle i|\rho|i\rangle$.}
\end{figure}

This figure shows that the population along the line of $m=-8$
decays at $(-8,5)$ in Fig.\ 2 and
the populations along the line $m=-9$ and $-10$ decay 
at $(-9,4)$ and $(-10,3)$, respectively.
These points are shown by squares in Fig.\ 2.

%As far as we see the energy gap $\Delta E$ along a line, 
%the energy gap increases monotonically. However 
If the system has an equilibrium distribution
in the initial state, the population distributes on levels. 
The population at each line decays at an avoided level 
crossing point where the LZS transition probability
\begin{equation}
p=1-\exp\left[ -{\pi(\Delta E)^2\over 2 c |m - m'|} \right]
\label{p}
\end{equation}
has an appreciable value. 
Here $\Delta E$ is the energy gap at $(m,m')$.
In Table I the transition probabilities are listed. 
For example the transition probability for $c=0.0001$
at $(m, m')=(-8,5)$ (see Fig. 3)
is 0.913, while at the point $(-8,6)$ it is $0.07$.
Thus most of the population of the line of $m=-8$ decays at 
$(-8,5)$. The population of the line of $m=-9$ decays very little until  
$(-9,4)$ because the transition probabilities at 
$(-9,6)$ and $(-9,5)$ are very small, i.e., 0.0006 and 0.04, 
respectively.
The population of the line of $m=-10$ decays at the point $(-10,3)$.

%At the points denoted by squares in Fig.\ 2, the transition probabilities
%are large (i.e., $p_i \simeq 1.0$). However the population coming to the point
%(e.g., the population on the line $m=-8$ for the point $(-8,5)$) decays
%in the previous crossing point (that is, $(-8,6)$) and the relaxation
%at the squares are very little because of little population.

As we found here, 
the parity effect simply comes from the structure of the energy levels.
The structure of energy levels in Fig.\ 2 is inherent to the 
systems of the uniaxial Hamiltonian (\ref{ham}) regardless of the form $Q$
and we expect that the parity effect is observed generally in uniaxial
magnets.

Here it should be noted that the relevant sequence of decays (in the above
case $H=0.3, 0.5$, and $0.7$) 
depends on the sweeping rate $c$. If $c$ decreases,
then the transition probabilities increase. Thus the populations 
on the lines decay at the circled avoided level crossing points 
before reaching the points of squares.
Thus in this case relaxation is enhanced at $ H=0.2, 0.4$, and $0.6$
instead of $ H=0.3, 0.5$, and $0.7$ in the case of $c=0.0001$.
In Fig.\ 4, we plot the time evolution of the magnetization process 
in case of 20 times slower sweeping rate $c=0.000005$. 
In Fig.\ 5, we show the time evolution of probabilities at levels, where
we actually find large decreases of the population at $H=0.2, 0.4$, and $0.6$.
Furthermore, in case of much slower 
sweeping rate $c=0.00000005$, we expect
that the transitions occur at the points of triangles in Fig.\ 2 (see also Table I).
We do not demonstrate it because it takes $100$ times longer simulation time. 
This sweeping rate dependence is also a general property of 
uniaxial magnets and 
we expect the shift of the sequence to be also found in experiments.

\begin{figure}
\noindent
\epsfxsize=9.0cm \epsfysize=6.0cm \epsfbox{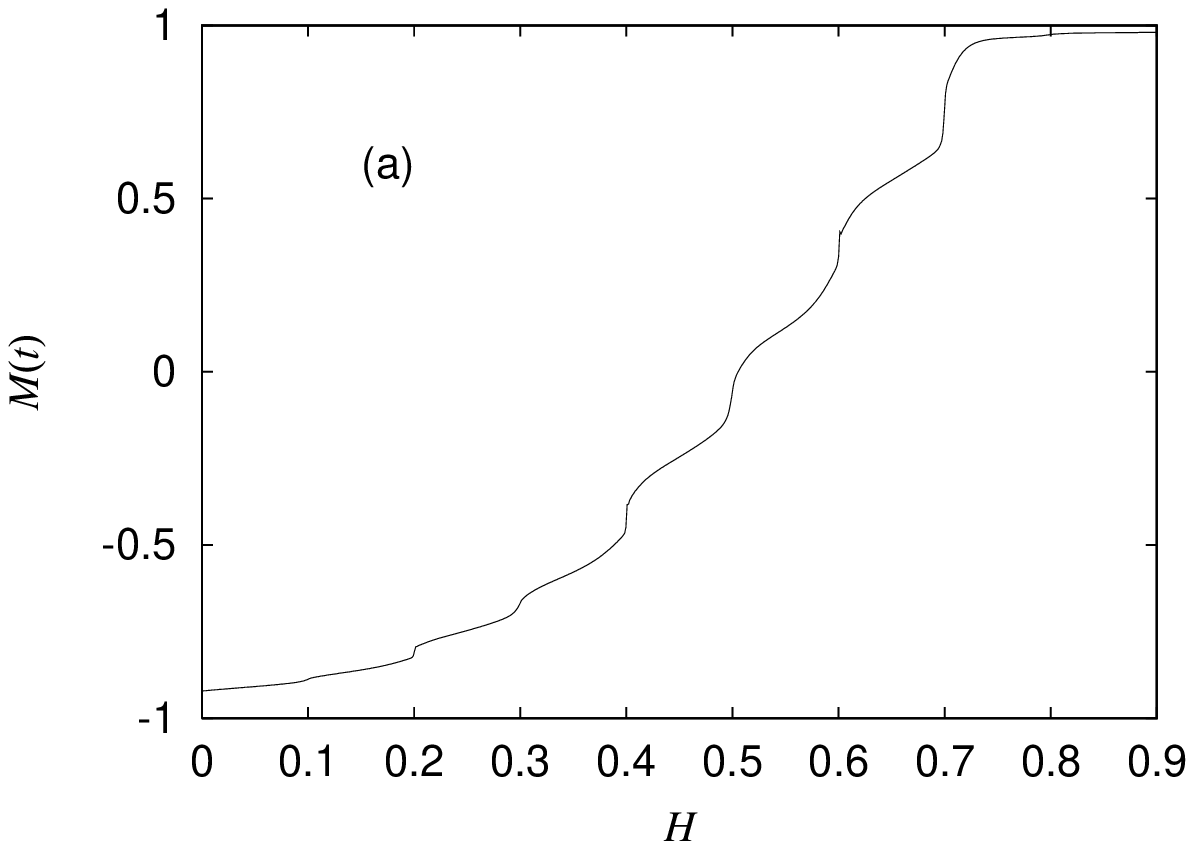}\\
\epsfxsize=9.0cm \epsfysize=6.0cm \epsfbox{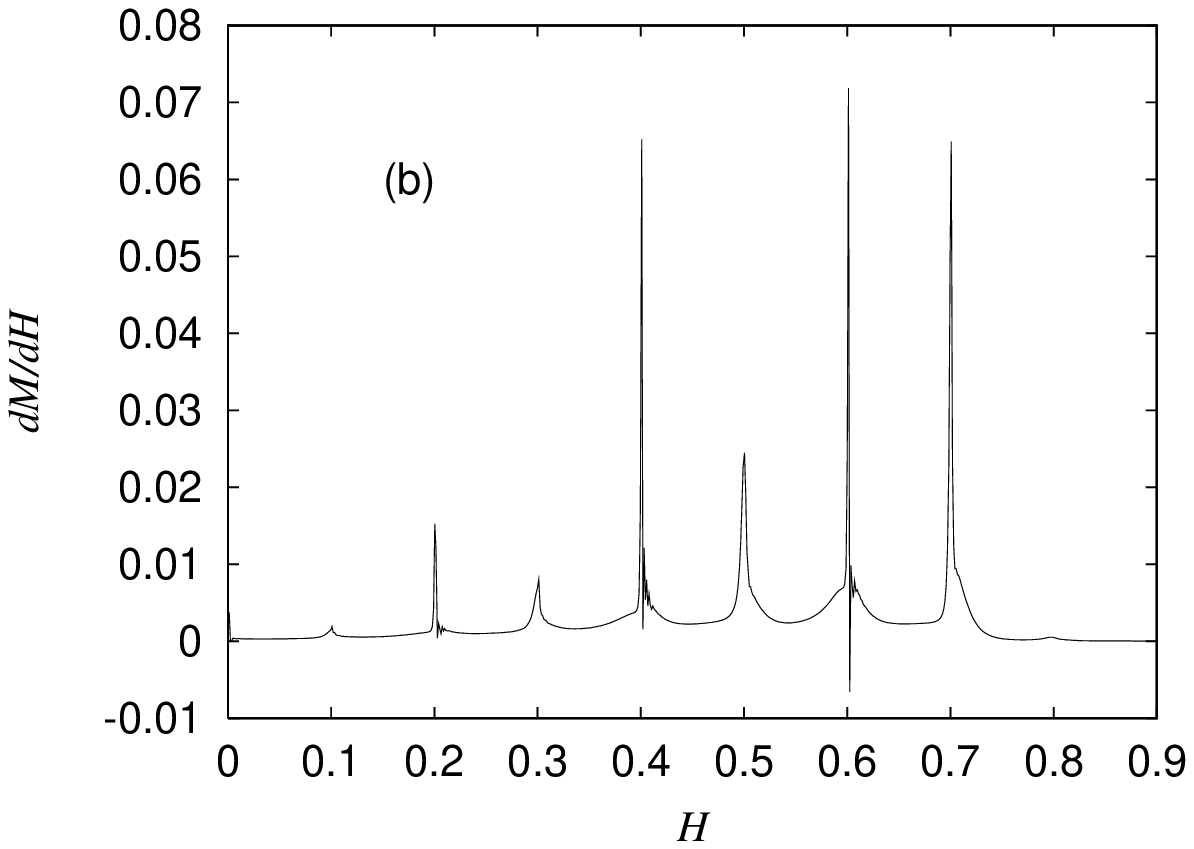}\\
\caption{(a) Magnetization process and (b) $dM/dH$ for $c=0.000005$.}
\end{figure}
\begin{figure}
\noindent
\epsfxsize=9.0cm \epsfysize=6.0cm \epsfbox{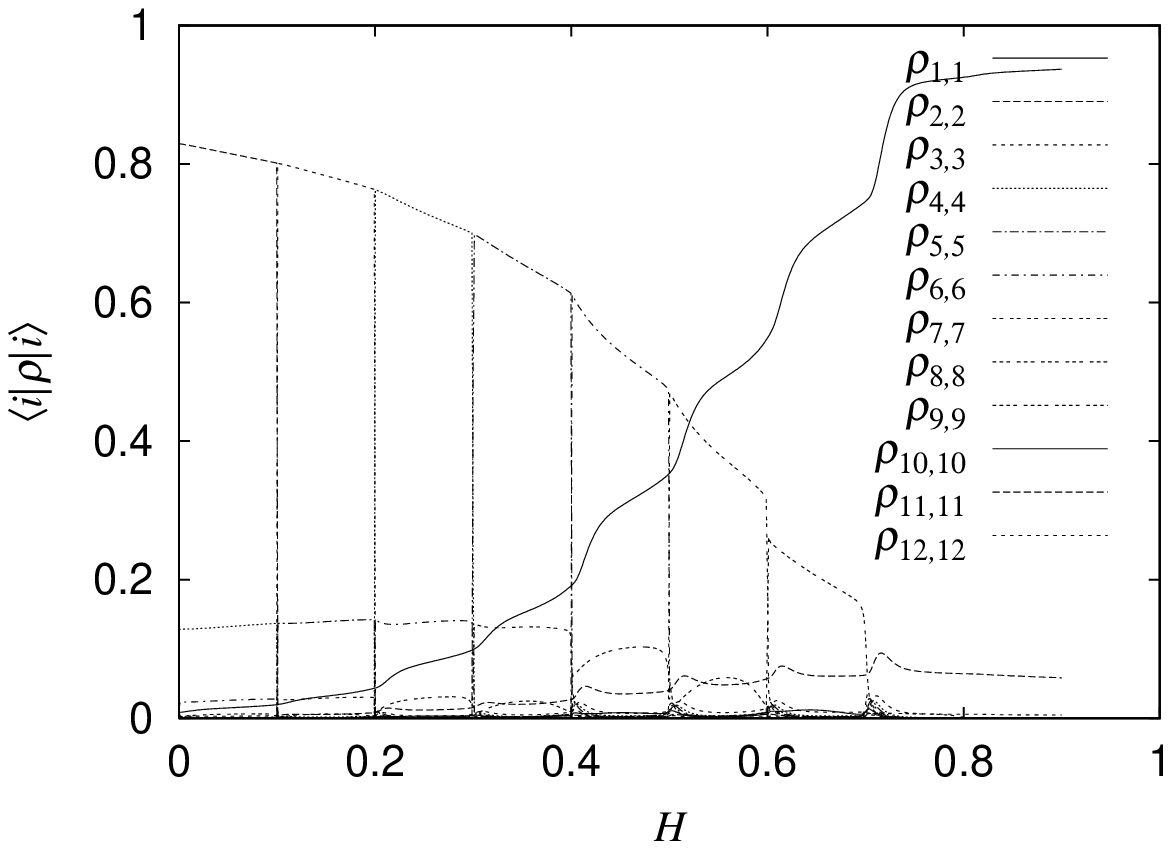}
\caption{Time evolution of the population $\langle i|\rho|i\rangle$.}
\end{figure}

Finally we would like to point out a strange property of the system when the Hamiltonian 
(\ref{ham}) includes the interaction 
\begin{equation}
Q=C[(S^+)^4+(S^-)^4]
\label{qterm}
\end{equation}
which has been discussed in literatures\cite{BB1,FRVGS98}.
As has been pointed out by Wernsdorfer and Sessoli\cite{Fe2},
in the present model
energy gaps at the avoided level crossing point change nonmonotonically 
with the value of $C$ ($\Gamma$ is fixed) and even gapless points exist, 
which causes irregular 
behavior of resonant tunneling and simple parity effect is disturbed.
%The same behavior is also found in the present system (\ref{ham}). 
%In Fig.\ 6, the energy gap at the avoided level crossing point $(-8,6)$ 
%is plotted as a
%function of the strength of the fourth-order term (\ref{qterm}),
The reason why gapless points appear is not clear at this moment,
which would be an interesting problem.
%\begin{figure}
%\noindent
%\epsfxsize=9.0cm \epsfysize=6.0cm \epsfbox{gap_045.eps}
%\caption{Dependence of the energy gap on $C$. Here the transverse field 
%$\Gamma$ is fixed to be $-0.45$.}
%\end{figure}

We would like to thank B. Barbara, H. De Raedt and W. Wernsdorfer for their valuable
discussions. The present work is partially supported by the Grant-in-Aid
from the Ministry of Education.

%\end{document}
\onecolumn
\begin{table}
\begin{tabular}{|c|c|c|ccc|}
\hline
   $H$ & $\Delta E$ & $(m,m')$ &$p(c=0.00000005)$&$ p(c=0.000005)$ &$p(c=0.0001)$\\
\hline
  0.000&  0.1554D$-$05& $(-10,10)$ &  0.3795D$-$05&  0.3795D$-$07&  0.1897D$-$08\\
  0.000&  0.1402D$-$05& $(-9,9)$   &  0.3432D$-$05&  0.3432D$-$07&  0.1716D$-$08\\
  0.000&  0.8514D$-$04& $(-8,8)$   &  0.1413D$-$01&  0.1423D$-$03&  0.7116D$-$05\\
  0.000&  0.9031D$-$02& $(-7,7)$   &  0.1000D+01&  0.8396D+00&  0.8745D$-$01\\
  0.100&  0.1466D$-$05& $(-10,9)$  &  0.3555D$-$05&  0.3555D$-$07&  0.1778D$-$08\\
  0.100&  0.5161D$-$05& $(-9,8)$   &  0.4923D$-$04&  0.4923D$-$06&  0.2461D$-$07\\
  0.100&  0.9741D$-$03& $(-8,7)$   &  0.8629D+00&  0.1968D$-$01&  0.9932D$-$03\\
  0.101&  0.5483D$-$01& $(-7,6)$   &  0.1000D+01&  0.1000D+01&  0.9735D+00\\
  0.200&  0.1391D$-$05& $(-10,8)$  &  0.3375D$-$05&  0.3375D$-$07&  0.1688D$-$08\\
  0.200&  0.7220D$-$04& $(-9,7)$   &  0.1018D$-$01&  0.1023D$-$03&  0.5118D$-$05\\
  0.200&  0.8156D$-$02& $(-8,6)$   &  0.1000D+01&  0.7752D+00&  0.7191D$-$01\\
  0.300&  0.3558D$-$05& $(-10,7)$  &  0.2340D$-$04&  0.2340D$-$06&  0.1170D$-$07\\
  0.300&  0.7418D$-$03& $(-9,6)$   &  0.6841D+00&  0.1146D$-$01&  0.5760D$-$03\\
  0.302&  0.4497D$-$01& $(-8,5)$   &  0.1000D+01&  0.1000D+01&  0.9131D+00\\
  0.400&  0.4119D$-$04& $(-10,6)$  &  0.3325D$-$02&  0.3331D$-$04&  0.1666D$-$05\\
  0.400&  0.5846D$-$02& $(-9,5)$   &  0.1000D+01&  0.5356D+00&  0.3762D$-$01\\
  0.500&  0.3846D$-$03& $(-10,5)$  &  0.2664D+00&  0.3093D$-$02&  0.1549D$-$03\\
  0.502&  0.2909D$-$01& $(-9,4)$   &  0.1000D+01&  0.1000D+01&  0.6404D+00\\
  0.600&  0.2938D$-$02& $(-10,4)$  &  0.1000D+01&  0.1761D+00&  0.9639D$-$02\\
  0.702&  0.6935D$-$01& $(-10,3)$  &  0.1000D+01&  0.1000D+01&  0.9970D+00\\
\hline
\end{tabular}
\caption{Energy gaps at avoided level crossings and LZS probabilities}
\end{table}
\end{document}